\documentclass[pre,twocolumn,showpacs,amsmath,superscriptaddress,amssymb]{revtex4}
\usepackage[dvips]{graphicx}
\usepackage{dcolumn}

\begin{document}
\title{High pressure dynamics of hydrated protein in bio-protective trehalose environment}
\author{S.O. Diallo}
\email{omardiallos@ornl.gov}
\affiliation{Quantum Condensed Matter Division,  Oak Ridge National Laboratory, Oak Ridge, Tennessee 37831, USA}
\author{Q. Zhang}
\affiliation{Biology and Soft Matter Division, Oak Ridge National Laboratory, Oak Ridge, Tennessee 37831, USA    }
\author{H. O'Neill}
\affiliation{Biology and Soft Matter Division, Oak Ridge National Laboratory, Oak Ridge, Tennessee 37831, USA    }
\author{E. Mamontov}
\affiliation{Chemical and Engineering Science Division,  Oak Ridge National Laboratory, Oak Ridge, Tennessee 37831, USA}


\begin{abstract}
We present a pressure dependence study of the dynamics of  lysozyme protein powder immersed in deuterated $\alpha$,$\alpha$-trehalose environment via quasi-elastic neutron scattering (QENS). The goal is to  assess the baro-protective benefits of trehalose on bio-molecules by comparing the findings with those of a trehalose-free reference study. While the mean-square displacement of the trehalose-free protein (hydrated to $d_{D_2O}\simeq$40 w\%) as a whole,  is reduced by increasing pressure, the actual observable  relaxation dynamics in the pico-(ps) to nano-seconds (ns) time range remains largely unaffected by pressure - up to the maximum investigated pressure of 2.78(2) Kbar. Our observation is independent of whether or not the protein is mixed with the deuterated sugar. This suggests  that the hydrated protein's conformational states at atmospheric pressure remain unaltered by hydrostatic pressures, below 2.78 Kbar. We also found the QENS response  to be totally recoverable after ambient pressure conditions are restored. Circular dichroism and neutron diffraction measurements confirm that the protein structural integrity is conserved and remains intact, after pressure is released. We observe however a clear narrowing of the quasi-elastic neutron (QENS) response as the temperature is decreased from 290 K to 230 K in both cases,  which we parametrize using the Kohlrausch-Williams-Watts (KWW) stretched exponential model.  Only the fraction of protons that are immobile on the accessible time window of the instrument, referred to as the elastic incoherent structure factor or (EISF) is observably sensitive to pressure, increasing only marginally but systematically with increasing pressure. \end{abstract}
\maketitle

\date{\today}

\section{Introduction}\label{sec0}
Understanding the mechanism by which organisms survive under extreme environments such as excessive heat and/or dehydration in arid  or hot regions, unusual cold in the arctic, or elevated pressures at the bottom of the oceans, is a topic of chief scientific relevance in biology, and physiology \cite{Clegg:01,Feofilova:03,Richards:02}.  While this survival ability  has long been known to be due to the presence of  non-reducing disacharides, such as trehalose, in certain living cells and plants, the underlying process by which these sugars stabilize biological systems is far from being fully understood. Among its bio-protective benefits, trehalose is known for example to help preserve the structural integrity in halophiles and cyanobacteria \cite{Crowe:84}, to serve as a carbon source or as a compatible solute for relieving high osmotic stresses in prokaryotes such as {\it Escherichia} coli during bio-synthesis \cite{Arguelles:00}. For these reasons, trehalose is also commonly used in industry for preserving food, vaccines, and cosmetic products \cite{Richards:02}.

 To date, two main scenarios have been proposed to explain how trehalose is able to serve as a good bio-protective agent, with some experimental evidence supporting both. Each proposal has only been able to explain a portion of the mechanism. Green and Angell \cite{Green:89} for example have related the resistance to extreme temperatures to the high glass transition temperature of trehalose with respect to that of pure water, which allows for a protective vitrified sugar shield around bio-molecules.  Crowe and collaborators \cite{Crowe:84}, on the other hand, associated the resistance to drought to the ability of trehalose to establish strong hydrogen-bond-based interactions with the polar groups of bio-systems. In this latter scenario, trehalose is able to \lq {\it replace}' water near biological surfaces, thereby preserving the hydrogen bond network even in the absence of water. The neutron diffraction measurements of Branca {\it et al.} \cite{Branca:02,Branca:02a} reveal a strong distortion of the peaks linked to the hydrogen bonded network in the partial radial distribution functions for all disaccharides, and for trehalose in particular, consistent with this {\it replacement} theory.   Various spectroscopy techniques \cite{Faraone:01,Koper:08,Lelong:07, Magazu:01,Magazu:06,Magazu:07,Affouard:05, Chryssomallis:81} and molecular dynamics simulations \cite{Pereira:04,Lerbret:07,Magazu:10b,Revsbech:12} have consistently confirmed the slowing down of water molecules that are immersed in a trehalose environment at normal atmospheric pressure. This reduction in mobility is hypothesized to be linked to the formation of a more crystalline structure (a glassy shell that protects biological cells) as a result of hydrogen binding between the water and the trehalose molecules, consistent with Green and Angel's {\it glassy shell} proposal \cite{Green:89}.  While these represent important developments in the field, much work remains to be done before a full and complete picture can emerge regarding the mechanism by which trehalose facilities bio-protection.  An outstanding pertinent question in biology, is whether or not trehalose offers baro-protective benefits for bio-species and if so, to what extent and how. 
 
 Pressure is  a clean thermodynamic tuning variable that can be used to define conformational states in protein \cite{Kauzmann:87,Kumar:08}.  By varying pressure, it is possible to explore the conformation space from the folded to the unfolded protein, as the partial molar volume is changed, and correlate these findings with the protein flexibility and various function. While, the dynamics of hydration water around proteins and that of proteins \cite{Gabel:05,Mamontov:10,Doster:88,Refaee:03,Benedetto:12}.  have been extensively investigated at ambient pressure, much less efforts have been devoted to high pressure research, owing primarily to technical limitations, which are now slowly being overcome. 

\begin{figure}
\includegraphics[width=1.00\linewidth]{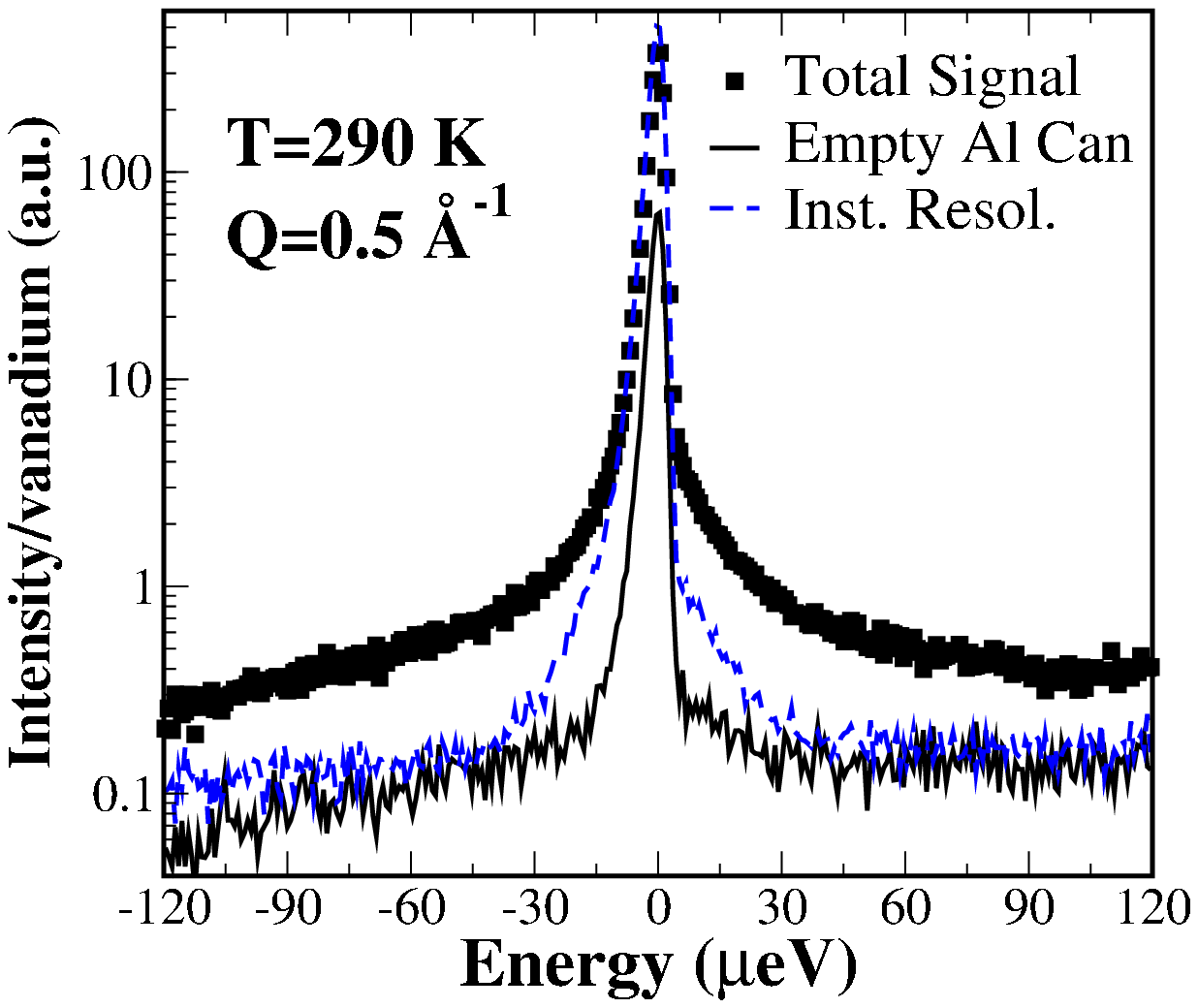}
\caption{Representative raw quasi-elastic neutron (QENS) response collected on the BASIS spectrometer at temperature $T=$290 K and wavevector $Q$=0.5 {\AA}$^{-1}$. The black circles show  the total signal from the D$_2$O-hydrated lysozyme powder in the Al container. The black solid line represents the signal from the empty high pressure Al cell and the blue dashed line shows the instrument resolution function, measured at 100 K using the same D$_2$O-hydrated sample.}
\label{fig.1}
\end{figure}

\begin{figure}
\includegraphics[width=1.00\linewidth]{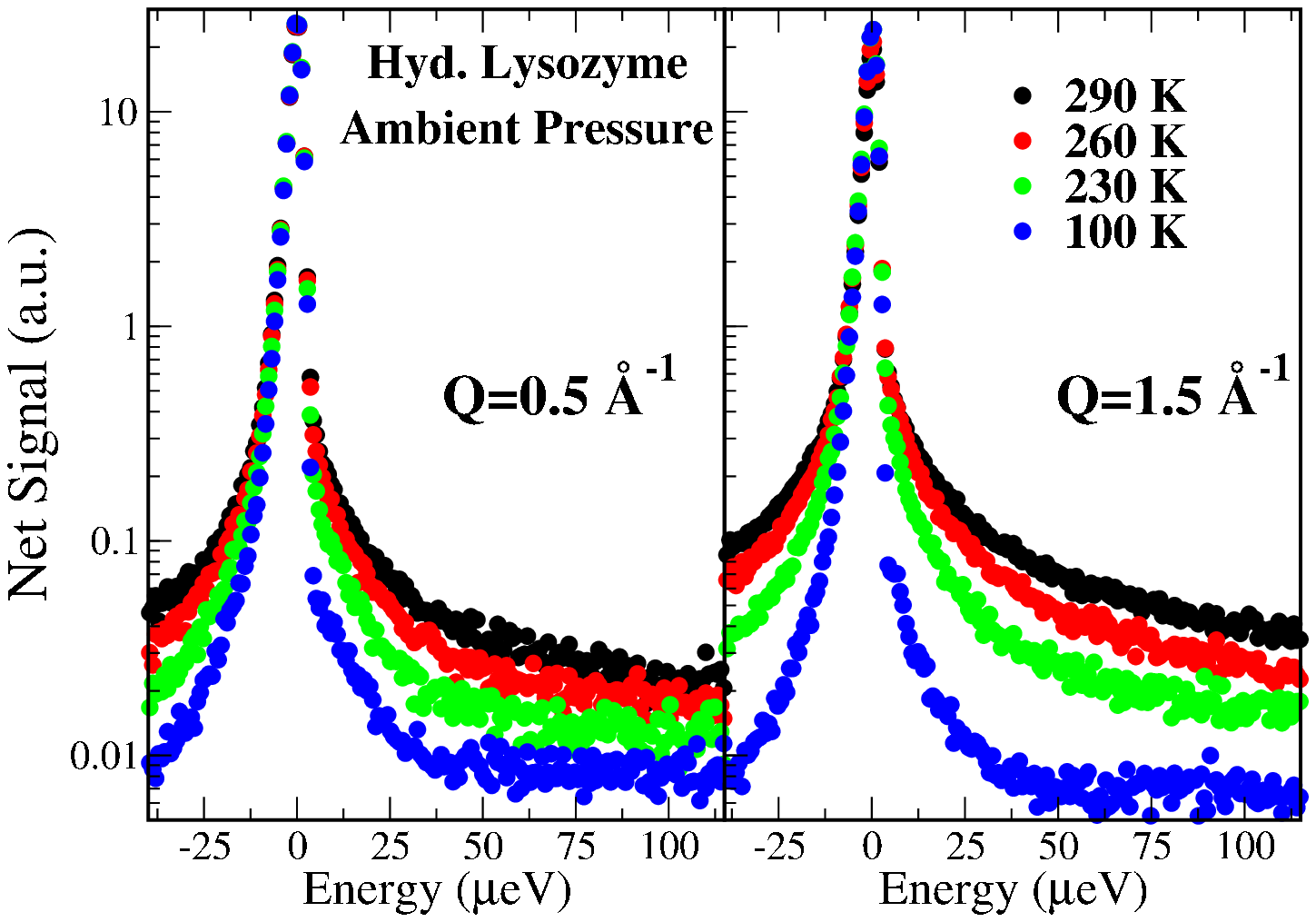}
\caption{Temperature dependence of the net QENS response from D$_2$O-hydrated lysozyme after subtraction of container contribution and  vanadium normalization at $Q$=0.5 {\AA}$^{-1}$ (left panel) and $Q$=1.5 {\AA}$^{-1}$ (right panel). Data at 100 K was used as a reference resolution function to determine the characteristic relaxation parameters at the higher temperatures.}
\label{fig.2}
\end{figure}

A technique of choice for studying protein dynamics is Quasi-Elastic Neutron Scattering (QENS) because it provides direct and unique information on the internal diffusive modes of hydrogen atoms in protein and their spatial correlations, from which the global conformational fluctuations of the protein can be inferred \cite{Zanotti:99,Zaccai:00a,Gabel:02}.  Several studies of various globular protein and of trehalose-water compounds at ambient pressure have already been reported 
\cite{Bellissent-Funel:96,Faraone:01,Koper:08,Lelong:07,Garcia-Sakai:13,Magazu:10,Affouard:05}. In a recent comprehensive study using both X-ray and neutrons, Ortore {\it et. al} \cite{Ortore:09} have simultaneously investigated the effect of high pressure on the structure and dynamics of lysozyme solution, up to about 1.5 Kbar. While they observe significant modifications in the protein-protein interaction potential just above 0.6 Kbar, they found no dramatic change in the protein globular structure with pressure.  They found a strong  correlation between the protein local dynamics and the water solvent, in agreement with an earlier QENS work on lysozyme in solution by Fillabozzi {\it et. al} \cite{Fillabozzi:05} in which the pressure dependence of the dynamics of lysozyme in solution was examined up  to $\sim$ 1.2 Kbar.

 We here present  high precision QENS measurements of D$_2$O-hydrated hen-egg-white lysozyme mixed with deuterated $\alpha$,$\alpha$-trehalose. Our aim is to evaluate how trehalose affects the dynamics of biological systems, when subjected to elevated pressures. We find that beyond a slow but systematic decrease of the fraction of immobile hydrogens in the protein (methyl and non-methyls groups static on the accessible time window on the neutron instrument) with pressure, there is no significant impact on the characteristic relaxation times in the nano- to pico-seconds range at all temperatures investigated, up to 2.78 Kbar. Interestingly, we find the QENS response and characteristic relaxations to be recoverable after ambient conditions are restored. These results indicate that the slow dynamics of hydrated lysozyme do not change with increasing hydrostatic pressures, in agreement with previous QENS reports \cite{Ortore:09,Fillabozzi:05} and  MD simulations \cite{Calandrini:08}.
 
 This article is organized as follows: technical aspects; primarily sample information and neutron measurements are presented in Sec. \ref{sec:exp}. Sec. \ref{sec:dan} discusses the data and fitting methods, followed by Sec. \ref{sec:res} where the main results are presented.  A summary is then presented in Sec.\ref{sec:sum}.

\section{Experimental Details} \label{sec:exp}
\subsection{Sample Preparation}
The lysozyme (L4919; 98\%purity) and $\alpha$,$\alpha$-trehalose deuterated samples were purchased from Sigma Aldrich, and Omicron respectively.   We first exchanged all labile hydrogen atoms for deuterium atoms by dissolving lysozyme in heavy water (D$_2$O), prior to lyophilization. The sample was then hydrated using isopiestic conditions by incubation in a sealed container containing respectively 99.9\% of D$_2$O. The level of hydration was controlled by varying the incubation time. The final hydration level $d_{D_2O}$ was determined by the relative change in the sample weight following humidity exposure, yielding a $d_{D_2O}\simeq$40\%. The hydrated batch was then used to prepare two samples for the neutron scattering measurements, one reference sample containing the hydrated lysozyme alone, and another one mixed with trehalose  (1 g of protein/ 1 g of sugar).  Approximately $\sim$ 150 mg of protein powder was used to prepare each sample, and loaded into a specially designed high pressure Al cell. We abandoned our original attempt to directly hydrate the dried lysozyme-trehalose mixture because it lead to a gel-like compound, which was quite different in texture and color than that obtained from hydrating the sugar-free sample. It is likely that in the presence of the sugar,  hydration preferentially starts with the sugar before wetting the protein. In this case, uniform protein hydration is hard to accomplish. The alternative approach we adopted above ensures that the protein gets hydrated to the desired level, before it gets in contact with the trehalose. This  yields comparable hydration levels in both of our samples.  

\begin{figure}
\includegraphics[width=1.00\linewidth]{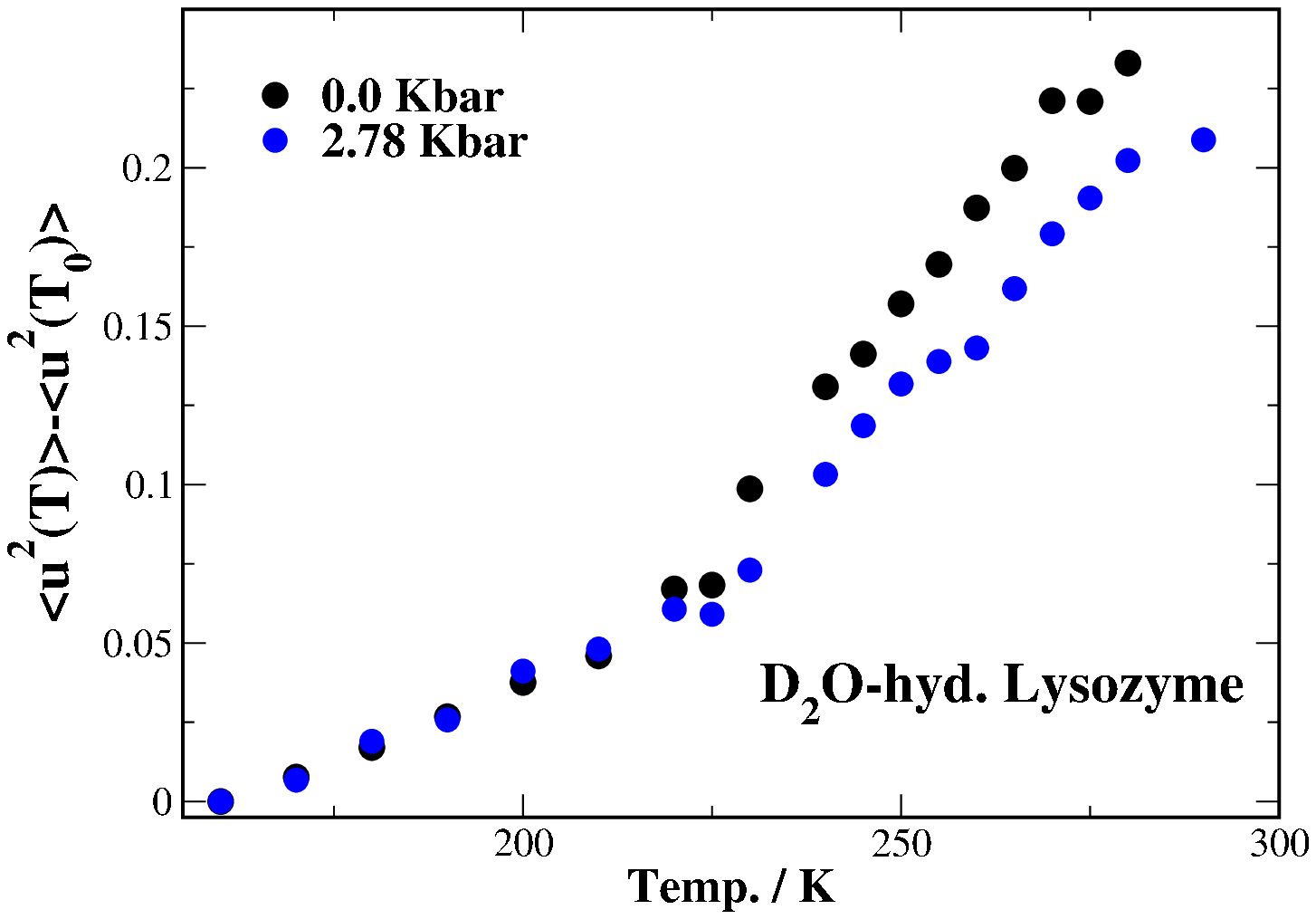}
\caption{Evolution of the mean square displacement (msd) of hydrogens with motion faster than $\sim$1 nanosecond, in D$_2$O-hydrated lysozyme powder with pressure. The harmonic behavior observed at low temperatures changes slope around 220 K. This anharmonicity at the high temperature goes down with increasing pressure. The reference msd value was inferred from the data at  $T_0=$150 K.}
\label{fig.3}
\end{figure}

\subsection{Elastic and Quasi-Elastic Neutron Scattering}
The neutron scattering measurements were performed on the backscattering spectrometer (BASIS) at the 1.4 MW Spallation Neutron Source, Oak Ridge National Laboratory (ORNL), USA \cite{Mamontov:11}, which has an energy resolution of 1.75 $\mu$eV (Half-Width at Half-Maximum) at the elastic line, and spans  a wide range of momentum transfer and energy transfer, respectively 0.3 $<Q<$ 1.9 {\AA}$^{-1}$, and  -120 $<\omega<$ 120  $\mu$eV. The useful QENS data were however analyzed over $Q$ in the range 0.5 $\le Q\le$1.5  {\AA}$^{-1}$. This was necessary to avoid coherent contribution from the protein at low and high $Q$'s. We used a helium gas panel with an intensifier to increase the pressure inside a specially designed Al cell, sealing the cell for the rest of the experiment when the desired pressure is reached. This means that the high pressure measurements were all performed at constant volume ($V_0$), starting from 290 K  and following the thermodynamic curve $P=\left[nRT/V_{0}\right]$ on cooling.  Experiments were performed at ambient pressure  (0.00), 1.00, 1.58, and 2.78 Kbar at three temperatures: 290 K, 260 K and 230 K for both samples. The instrument resolution function was measured using the \lq frozen' sample at 100 K, where the proton mobility in the protein becomes resolution limited on the instrument. The empty can was also measured for data correction.  

\begin{figure}
\includegraphics[width=1.00\linewidth]{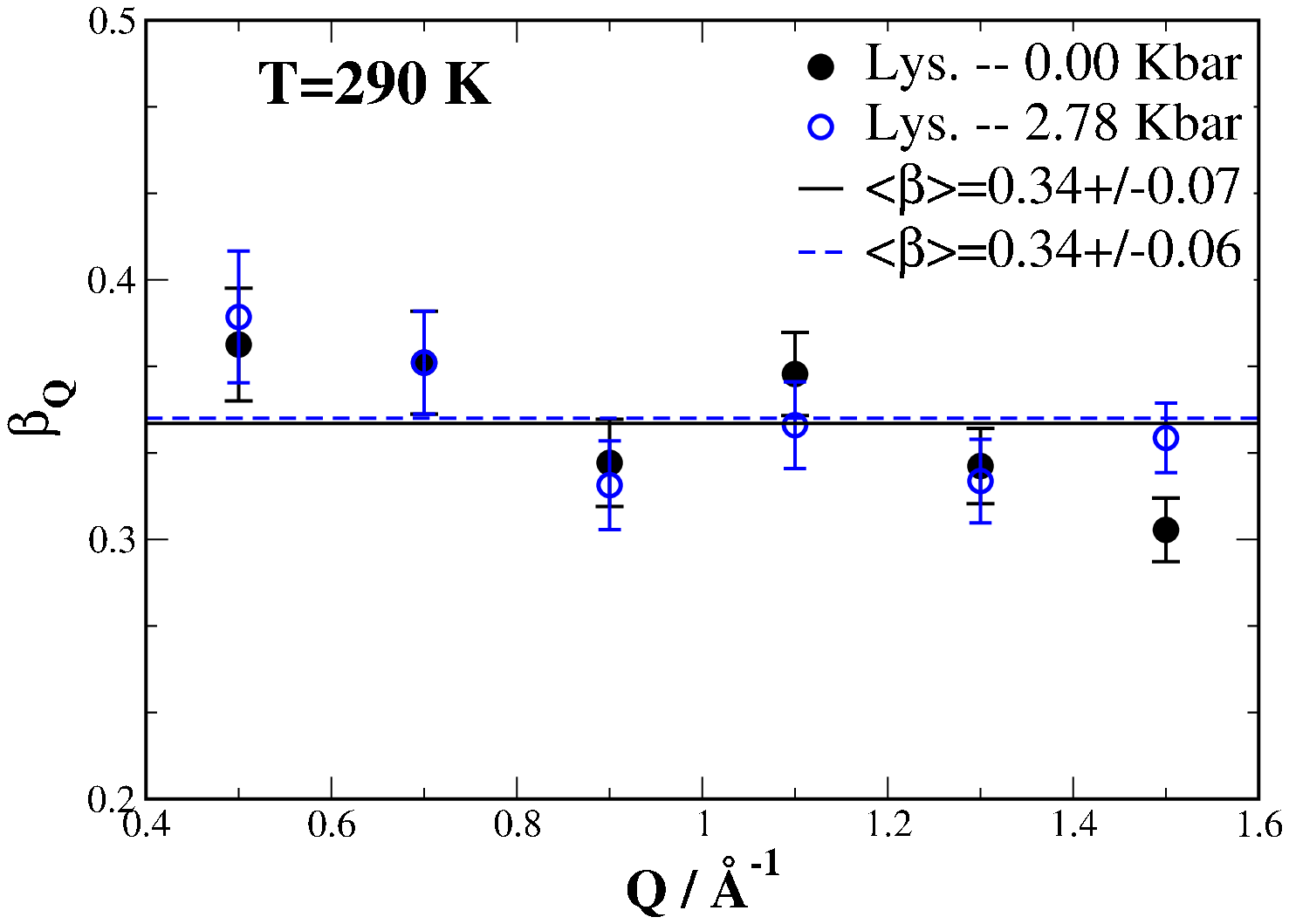}
\caption{Variation of the stretching exponent parameter $\beta_Q$ with momentum transfer $Q$ at temperature $T=$ 290 K. Solid circles are the observed values at ambient pressure, and the open circles at P=2.78 Kbar. The solid and dashed lines are the average values over all $Q$. }
\label{fig.4}
\end{figure}

\section{Data Analysis} \label{sec:dan}
Before any quantitative analysis, we first begin with a qualitative data inspection and simple comparison between the different spectra to look for trends and insure the observations are consistent with anticipated responses from the sample. We show as an example a representative raw spectra of hydrated lysozyme at $Q$=0.5 {\AA}$^{-1}$ and $T=$290 K in Fig. \ref{fig.1}. The instrument resolution function and the empty can data are also overlaid for comparison. The contributions from the empty can to the QENS signal are  largely limited to the elastic line and the linear background. We used a self-shielding factor of 1 to subtract the corresponding background.  Fig. \ref{fig.2} shows the temperature dependence of the net signal of lysozyme at ambient pressure after proper background correction at some selected $Q$ values. The data shows the temperature evolution of the QENS peak broadening with decreasing temperature, as would be expected for decreasing protein flexibility.

\begin{figure}
\includegraphics[width=1.00\linewidth]{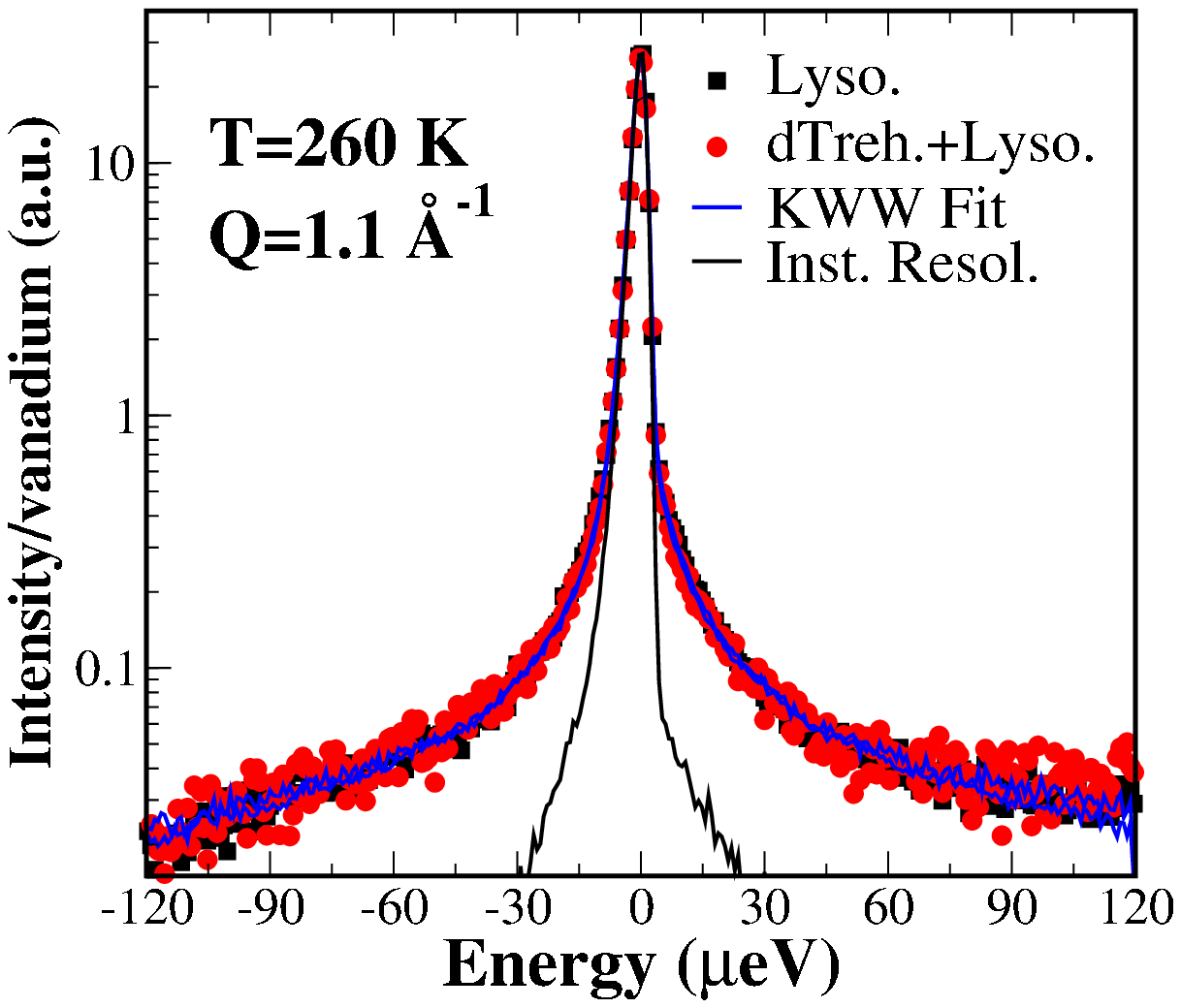}
\caption{Net observed signal (background subtracted) from the D$_2$O-hydrated protein compound without trehalose (black squares) and with trehalose (red circles). Solid black line shows the corresponding resolution function. The blue lines are the resulting fits to the data using a Kohlrausch-Williams-Watts (KWW) model, discussed in the text.}
\label{fig.5}
\end{figure}

\begin{figure}
\includegraphics[width=1.00\linewidth]{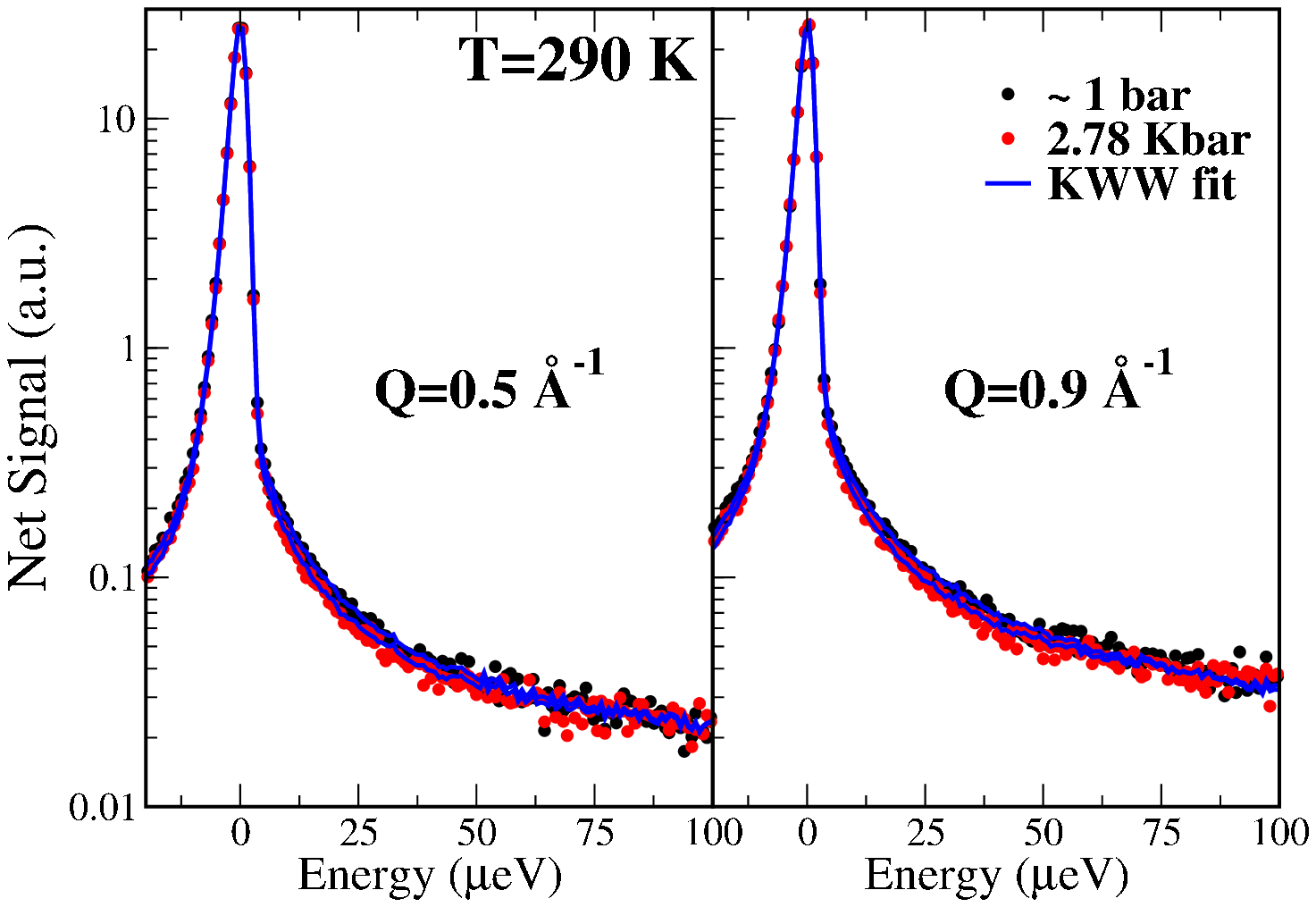}
\caption{Pressure dependence of the net response from D$_2$O-hydrated lysozyme at selected $Q$=0.5 {\AA}$^{-1}$ (left panel) and $Q$=0.9 {\AA}$^{-1}$ (right panel) at 290 K.  Black circles are the data at ambient pressure (1 bar) and red circles at 2.78 Kbar. The blue lines are the fits to the data using a Kohlrausch-Williams-Watts (KWW) model, as discussed in the text.}
\label{fig.6}
\end{figure}
Just prior to and immediately after the long QENS measurements, we performed diagnostic  \lq incoherent elastic intensity' scans on the D$_2$O-hydrated lysozyme sample, free of any trehalose, to look for differences in the global molecular fluctuations between 1 bar and 2.78 Kbar.  The elastically scattered neutrons  were recorded on heating from 150 K up to 290 K, in variable steps of 5 and 10 K, depending of the temperature range. The elastic intensity as a function of temperature was obtained by integrating the corresponding spectrum over a very small energy range comparable to that of the instrument resolution, for each $Q$. Assuming an isotropic flexibility in the motion of the hydrogens inside the protein, the mean square displacement $\langle u^2(T)\rangle$ (or MSD) can  be calculated from the elastic intensity $I_s(T)$ using the expression $\langle u^2(T)\rangle=-\frac{3}{Q^2}\ln\left[\frac{I_s(T)}{I_s(T_0)}\right]$ where $T_0$ represent the lowest measured temperature of 150 K.  Fig. \ref{fig.3} shows the derived MSD  as a function of temperature for the two pressures investigated. The data was calibrated relative to the ambient pressure MSD at 150 K.  As the temperature is increased,  $\langle u^2(T)\rangle$ increases harmonically up to about 220-230 K where it starts to increase more rapidly with increasing temperature. This deviation from harmonic motions \cite{Becker:03},  commonly found in bio-molecules (e.g: proteins, DNA, RNA...etc..) and is  generally referred to as the dynamical transition \cite{Doster:88,Chen:06,Liu:04,Roh:06}. From Fig. \ref{fig.3}, it is clear that anharmonic effects at 2.78 Kbar are less prominent than those at ambient pressure,  but they continue to be present.

To analyze the QENS data, we fitted each spectra independently using the DAVE software package \cite{Azuah:09}, according to the generic model $I(Q,\omega)$:
  \begin{eqnarray}
  I(Q,\omega)&=&N(Q) \Big[  EISF(Q)\delta(\omega)+(1-EISF(Q))\nonumber \\ 
              &&S_{m}(Q,\omega)\Big] \otimes R(Q,\omega) +B(Q,\omega) 
 \label{eq.IQ}
 \end{eqnarray}
\noindent where $N(Q)$ is an arbitrary scale factor, $EISF(Q)$ represents the population fraction of immobile protons or the elastic incoherent structure factor, $\delta(\omega)$ is a delta function centered around zero energy transfer, $B(Q,\omega)$ is a residual background term in the form $B(Q,\omega)=B_1+B_2(\omega+\omega_0)^{-1}$ (with $\omega_0$ fixed to the elastic energy  of 2080 $\mu$eV), $R(Q,\omega)$ is the resolution function, and $S_{m}(Q,\omega)$ is a model scattering function, which depends intrinsically on the sample. The internal dynamics of protein being far too complex to be represented by  a \lq standard' single Lorentzian function,  we used a stretched exponential function, also referred to as Kohlrausch-Williams-Watts (KWW) model \cite{Mansour:02,Takahara:05} to fit the data:
\begin{equation}
\label{eq.KWW}
S_m(Q,\omega)=\int_0^\infty dt e^{-[t/\tau(Q)]^{\beta_Q}}e^{i\omega t}.
\end{equation}
\noindent Here $\tau(Q)$ represent the relaxation time at a  particular $Q$, and $\beta_Q$ the stretching exponent, typically found to be 0 $<\beta_Q<1$ for systems with glassy behavior such as proteins. This model make physical sense and better account for the distribution in activation energy in the protein \cite{Roh:06}. Fig. \ref{fig.4} shows  the variation of  the observed stretching exponent $\beta_Q$ of lysozyme with momentum transfer $Q$ at $T=$ 290 K. It is clear that $\beta_Q$ has no significant dependence on $Q$, nor on pressure at 290 K, in agreement with previous work \cite{Bellissent-Funel:96,Nickels:13}. We thus kept $\beta_Q$ fixed to its average value of 0.34 in fitting the remainder of the data. This effectively reduces the free adjustable parameters to three: $N(Q)$, $EISF(Q)$ and $\tau(Q)$, excluding the background terms.

\section{Results}\label{sec:res}

\subsection{Protein Response and Influence of Trehalose}
Fig. \ref{fig.5} compares the ambient pressure QENS signal of lysozyme to that of the lysozyme-trehalose mixture at temperature $T=260$ K and $Q=1.1${ \AA}$^{-1}$. The lines represent the corresponding fits obtained with the KWW model. To the naked eye, there is no appreciable difference in the peak broadenings at this temperature. Further inspection of the data  at other temperatures and pressures yield essentially the same results. We thus proceeded to capturing the temperature dependence of the peak broadening as a function of temperature for all pressure investigated, since thermal effects are much more important.  Table \ref{tab1} summarizes some of the key findings, which we discuss below.

\begin{figure}
\includegraphics[width=1.00\linewidth]{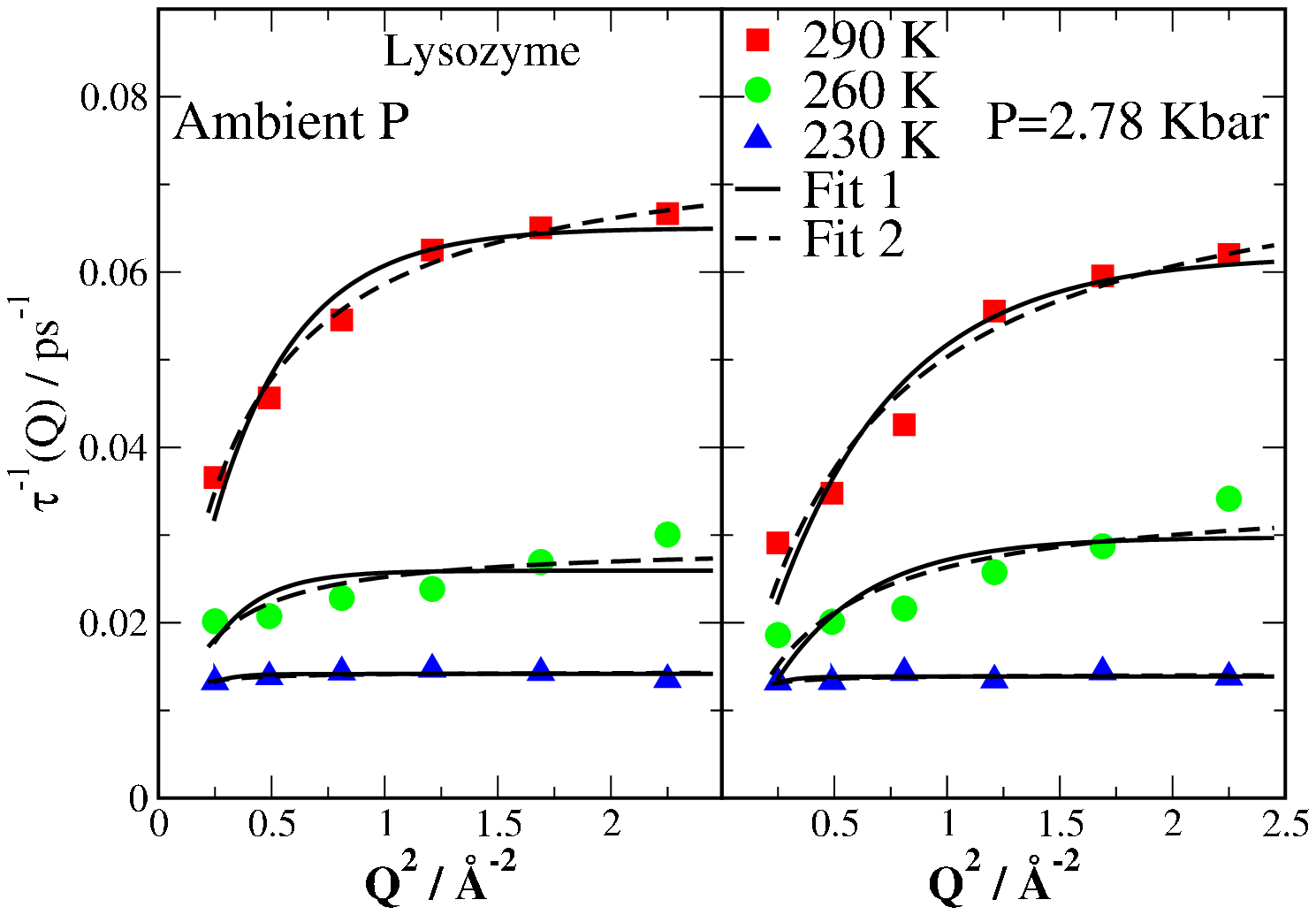}
\caption{Inverse $\tau_{Q}$ as a function of $Q^2$ for lysozyme at ambient pressure (left panel) and elevated pressure (right panel). Solid lines, denoted Fit 1,  are fits of Eq. \ref{eq.tauJump} to the observed values. Dashed lines (Fit 2) are fits of ${\tau^{-1}(Q)}=D_rQ^2/(1+D_r\tau_{0}Q^2)$ to the data. }
\label{fig.7}
\end{figure}

\begin{figure}
\includegraphics[width=1.00\linewidth]{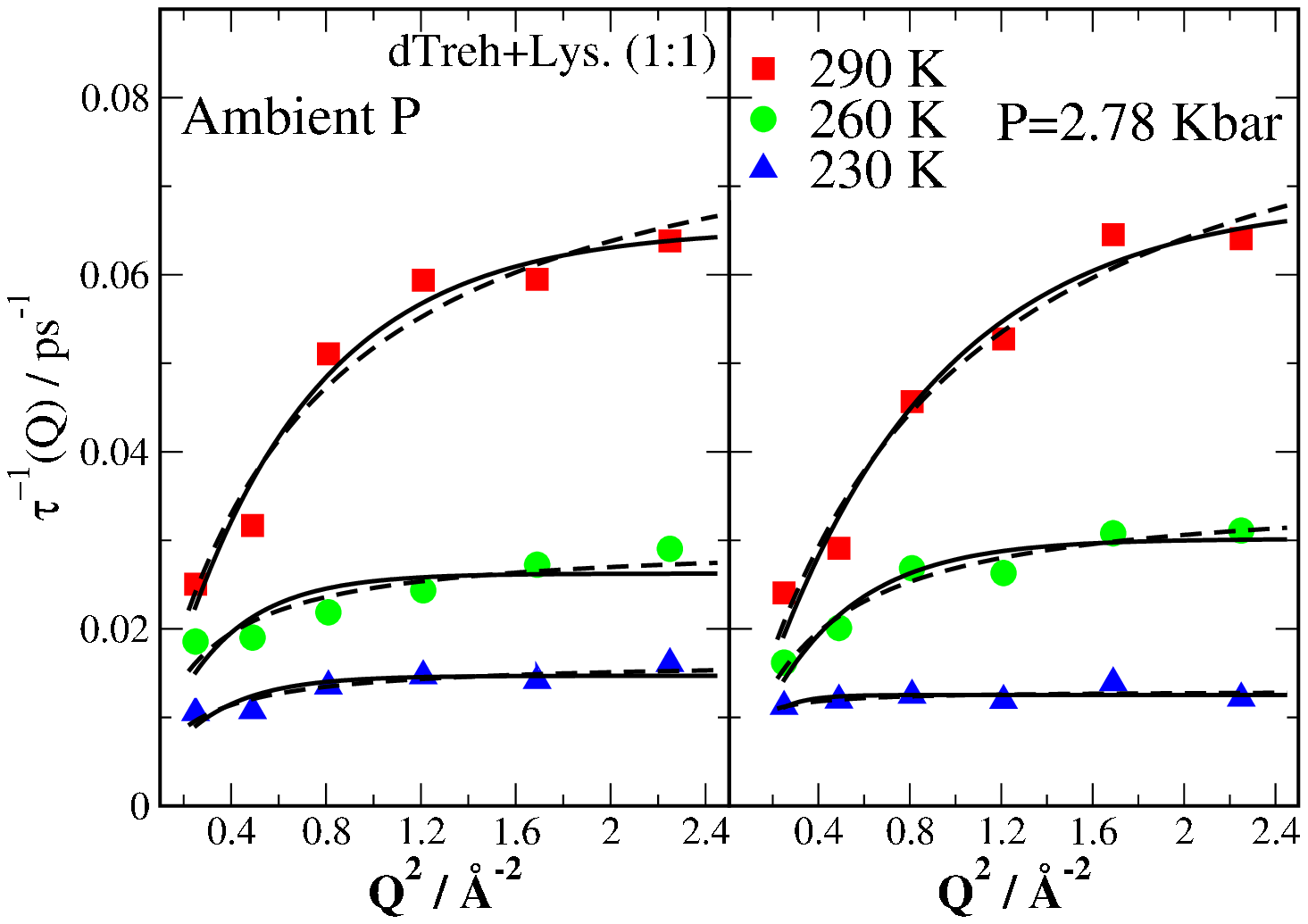}
\caption{Inverse $\tau_{Q}$ as a function of $Q^2$ for lysozyme and deuterated trehalose compound at selected pressures: ambient pressure (left panel) and elevated pressure (right panel). Labels are the same as in Fig. \ref{fig.7}}
\label{fig.8}
\end{figure}

\begin{table*}
\caption{EISF fit parameters and KWW residence times as a function of temperature and pressure.  The subscripts indicate the name of the sample, where $s$=$L$ is for lysozyme, and $s$=$LT$ indicates the lysozyme-trehalose compound.}
\label{tab1}
\begin{ruledtabular}
\begin{tabular}{||c|c||c|c||c|c||c|c||}
T/K &P/Kbar&$p_L$& $p_{LT}$&$f_L$& $f_{LT}$&$\tau_{r_{L}}$/ps& $\tau_{r_{LT}}$/ps\\
\hline
290	&0.00    & 0.489(1) & 0.570(5)	     & 0.820(1)        &	0.777(1) & 15.3&15.3\\
	&1.00   & 0.501(1)  & 0.550(1)	    & 0.830(1)          &	0.789(2)&14.3&15.7\\
	&1.58    & 0.510(1) & 0.585(1)	     &0.841(1)         &	0.811(1)&15.9&18.2\\
	&2.78   & 0.527(1)  & 0.570(1)	    & 0.865(1)          &	0.809(1) &16.2&14.5 \\
\hline
260	&0.00    & 0.512(1) &0.590(1)&0.870(1)& 0.880(1)	              &38.6&38.1\\
	&1.00   & 0.601(1)  &0.630(1)&0.890(1)& 0.878(2)	           &38.9&35.1\\
	&1.58    & 0.600(1) &0.630(1)&0.892(1)& 0.880(1)	           &43.6&41.8\\
	&2.78   & 0.630(1)  &0.627(2)&0.890(1) & 0.887(1)	    &33.6&33.2\\
\hline
230	&0.00    & 0.740(1) & 0.989(2)	     &  0.939(1)        &	0.869(1)&70.6&68.0\\
	&1.00   & 0.720(1)  & 0.877(1)	    & 0.940(1)         &	0.930(1)&71.7&53.0\\
	&1.58    & 0.731(1) & 0.779(1)	     &  0.940(2)       &	0.951(1)&74.2&69.0\\
	&2.78   & 0.729(2)  & 0.801(4)	    & 0.950(1)          &	0.940(1)&72.2&79.8\\
\end{tabular}
\end{ruledtabular}
\end{table*}

\subsection{Effects of Pressure} 

In this section, we evaluate how pressure affects the dynamics observed at atmospheric pressure. We begin first by investigating the trehalose-free sample with a comparative inspection of the ambient pressure data and that at 2.78 Kbar. Such a comparison  is illustrated by Fig. \ref{fig.6}, which shows the spectra collected at $T=$290 K for two selected $Q$'s. For clarity, data at intermediate pressure values (1 and 1.58 Kbar) have been omitted but lie well within the two pressure limits. As can be seen, there is no observable change in the QENS lineshape, as pressure is increased slowly from 1 bar to 2.78 Kbar, suggesting that the relaxation dynamics on the pico- to nano-second scale are not perturbed by hydrostatic pressure, below 3 Kbar.   We observe a very similar behavior with the data collected with the lysozyme immersed in trehalose. Nevertheless, we use Eqs. \ref{eq.IQ}, and \ref{eq.KWW} to document the relaxation parameter $\tau(Q)$, and the $EISF(Q)$ at all temperatures and pressures probed.   The variation  of the inverse of the relaxation time $\tau(Q)$ with $Q^2$ is displayed in Figs. \ref{fig.7} and \ref{fig.8}, for the two pressure limits: ambient and highest pressure. The relaxation dynamics at physiological temperature 290 K  depicts the strongest dependence with $Q$, and suggests a jump diffusion behavior. This coupling with $Q$ is reduced at 260 K, and barely noticeable at the lowest temperature of 230 K.  To parametrize $\tau(Q)$, we use the following model \cite{Hall:81}:  
\begin{equation}
\label{eq.tauJump}
\frac{1}{\tau(Q)}=\frac{1}{\tau_{r}} (1-e^{-D\tau_{r}Q^2})
\end{equation}
\noindent where $\tau_r$ is the residence time between jumps, and $D=\langle u^2\rangle/6\tau_r$ the diffusion coefficient. We found $D_r$ to be more reliably determined at 290 K, with $D_r$ in the range of 1.7-1.8 $\times$ 10 $^{-5}$ cm$^2$ s$^{-1}$ at ambient pressure and 0.9-1.0  $\times$ 10 $^{-5}$ cm$^2 $s$^{-1}$ at 2.78 Kbar in both samples. These values differ somewhat from the 2.5-3 $\times$ 10 $^{-5}$ cm$^2$ s$^{-1}$ recently reported by Ortore {\it et al} \cite{Ortore:09} using a localized diffusion in a sphere model for lysozyme in solution. This is not surprising since the determination of the local diffusion coefficient is known vary with the model. At 260 K, $D_r$ reduces down to $\sim$ 0.9-1.2 $\times$ 10 $^{-5}$ cm$^2$ s$^{-1}$ at zero pressure, and to 0.6-0.7 $\times$ 10 $^{-5}$ cm$^2$ s$^{-1}$ at the highest pressure. At 230 K however, the protein dynamics  becomes observably so small that they fall within the instrument resolution. In this case,  it becomes difficult to reliably resolve $D_r$, as can be anticipated from the $Q$-behavior of $\tau(Q)$ shown by Figs. \ref{fig.7}, and \ref{fig.8}. The fitted $D_r$ values at 230 K fluctuate nonetheless between 0.5-1.5 cm$^2$ s$^{-1}$ at all pressures investigated.

To check the influence of model on the diffusion coefficients, but also to improve the quality of the fits  obtained for $\tau(Q)$, specially in light of the poorer fits at the high $Q$ at 260 K, we have re-fitted the data using an alternate jump model ${\tau^{-1}(Q)}=D_rQ^2/(1+D_r\tau_{0}Q^2)$. These fits are displayed as dashed lines and denoted \lq Fit 2'  in Figs. \ref{fig.7}. With this alternate model, the observed diffusion coefficients at 290 K are indeed larger than those obtained with Eq. \ref{eq.tauJump}, yielding $D_r\simeq$2.5  $\times$ 10 $^{-5}$ cm$^2$ s $^{-1}$ at ambient pressure which decreases to 1.4 $\times$ 10 $^{-5}$ cm$^2$ s $^{-1}$ at 2.78 Kbar, and  much closer to those estimated by Ortore {\it et al} \cite{Ortore:09}. The corresponding values for the lysozyme+trehalose mixture are for instance 1.4 $\times$ 10 $^{-5}$ cm$^2$ s $^{-1}$ at ambient temperature and pressure conditions, and 1.1 $\times$ 10 $^{-5}$ cm$^2$ s $^{-1}$ at the highest pressure. The $\tau_0$ values obtained by this method are only marginally smaller than $\tau_r$  of the KWW model introduced above, by about 2-5 ps. The average relaxation time $\langle \tau_{av}\rangle$ which reflects the stretching effect of $\beta_Q$ can be computed if desired using the expression, $\langle \tau_{av}\rangle=\frac{\tau_{i}}{\beta_Q} \Gamma(\frac{1}{\beta_Q})$, where $\Gamma(x)$ is the Gamma function, and $\tau_i$ equals to $\tau_r$ or $\tau_0$. This effectively scales up the observed $\tau_i$ values by a factor $\sim5.57$ and reduces $D_r$ by the same factor,  since our $\beta_Q$ is fixed to 0.34. It will not affect their trends with temperature or pressure. Table \ref{tab1} shows the observed $\tau_r$ values, along with other parameters which are discussed below. The subscript $s$ in $\tau_{rs}$ indicates the name of the sample, where $s$=$L$ is for lysozyme, and $s$=$LT$ indicates the lysozyme-trehalose compound.  Again, the strongest influence on this particular parameter is not pressure but rather temperature, to which we return below. 

\begin{figure}
\includegraphics[width=1.00\linewidth]{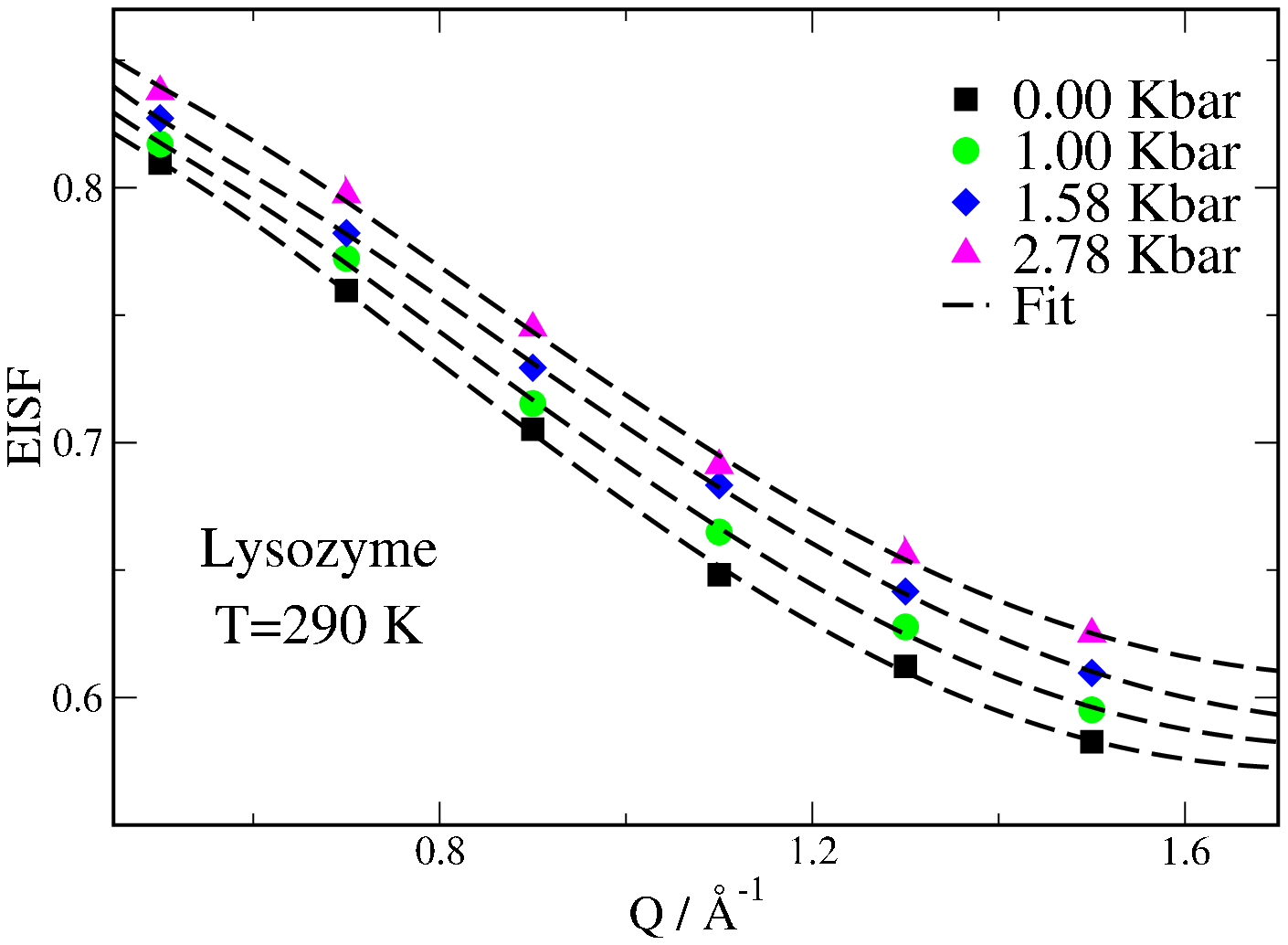}
\caption{Influence of pressure on the incoherent structure factor (EISF) of lysozyme at 290 K. Symbols represent results obtained respectively at ambient pressure (black squares), 1.00 Kbar (green circles), 1.58 Kbar (blue diamonds), and 2.78 Kbar (magenta triangles). Dashed lines are model fits to the data.}
\label{fig.9}
\end{figure}

\begin{figure}
\includegraphics[width=1.00\linewidth]{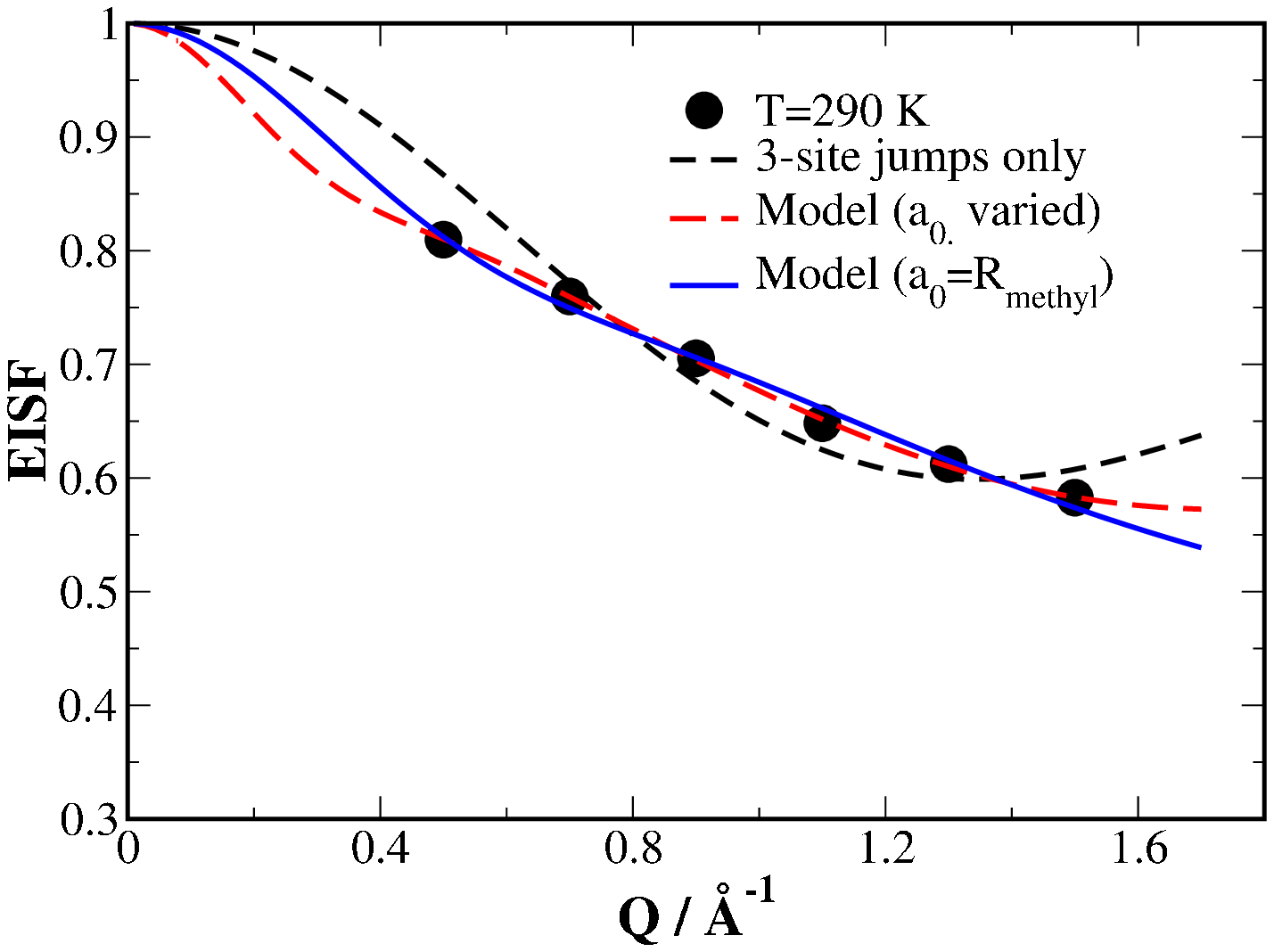}
\caption{Model fits to the elastic incoherent structure factor (EISF) determined from the D$_2$O-hydrated protein data at 290 K: black circles (experimental data); black short-dashed line (3-site jumps model); red long-dashed line (Eq.\ref{eq.eisf} with confining radius $a_0$ for methyls group allowed to vary);  blue solid line ( Eq. \ref{eq.eisf} with $a_0$ fixed to 1.1 {\AA}, as observed previously \cite{Roh:06}) .}
\label{fig.10}
\end{figure}

Based on the observations above,  we concluded that the QENS broadening ($\tau(Q)$ parameter) is not a relevant parameter for evaluating the effects of pressure on protein or for assessing the baro-protection of trehalose, as we originally have hoped for. We instead focused our attention to the only observable quantity that shows some systematic and discernible change with pressure; the $EISF(Q)$ introduced in Eq. \ref{eq.IQ}.  This parameter yields valuable information on the geometry of active motions observed on the QENS instrument \cite{Gabel:05}. Fig.\ref{fig.9} shows the $EISF(Q)$ parameter as a function of $Q$ for several pressures for lysozyme at 290 K. While the behavior with $Q$ appears to be the same, the magnitude of $EISF(Q)$ clearly increases with increasing pressure, suggesting that it is the population fraction - and not the relaxation times - of protons contributing to the different dynamical processes that gets affected by pressure. The dashed lines show fits to the data, based on the following coupled $EISF(Q)$ model that accounts for contributions from both methyl groups and non-methyl groups:

\begin{eqnarray}
\label{eq.eisf}
EISF(Q)&=&EISF_{meth.}(Q)\times EISF_{loc.}(Q) \\ \nonumber
	     &=&\left[p_s+\frac{1-p_s}{3}\left(1+2j_0(Q a_0\sqrt{3})\right)\right]\times \\
	     & &\left[f_s + (1-f_s)\left(\frac{3j_1(Q a_1)}{Q a_1}\right)^2 \right]\nonumber
\end{eqnarray}
\noindent where $p_s$ is the fraction of immobile observable protons associated with the methyl groups (3-fold jumps model), and $f_s$ that associated with non-methyl groups (generic localized dynamics). Corresponding confining radii for both groups are represented by $a_0$ and $a_1$, respectively.  The model above was necessary in the absence of non-hydrated samples (dry) data that would have otherwise allowed us to characterize the methyl groups alone \cite{Nickels:13}. Fig. \ref{fig.10} illustrates the models used to fit the $EISF(Q)$ obtained at 290 K for lysozyme, with Eq. \ref{eq.eisf}, and other variant fitting schemes. In fact, we used various models to fit a few selected $EISF(Q)$ before settling on that to use for the rest of the data. Specifically, we investigated a 3-sites jump model (assuming all $EISF$ arises from methyl groups only),  Eq. \ref{eq.eisf} with all parameters allowed to vary, and finally Eq. \ref{eq.eisf} with all but $a_0$ adjustable. In the later case, we set $a_0=1.1$ \AA, its reported value in the literature \cite{Roh:06}. Without this constraint, we get a somewhat larger $a_0$, with 1.3 $< a_0<$1.7 {\AA} at all temperatures.

\begin{figure}
\includegraphics[width=1.00\linewidth]{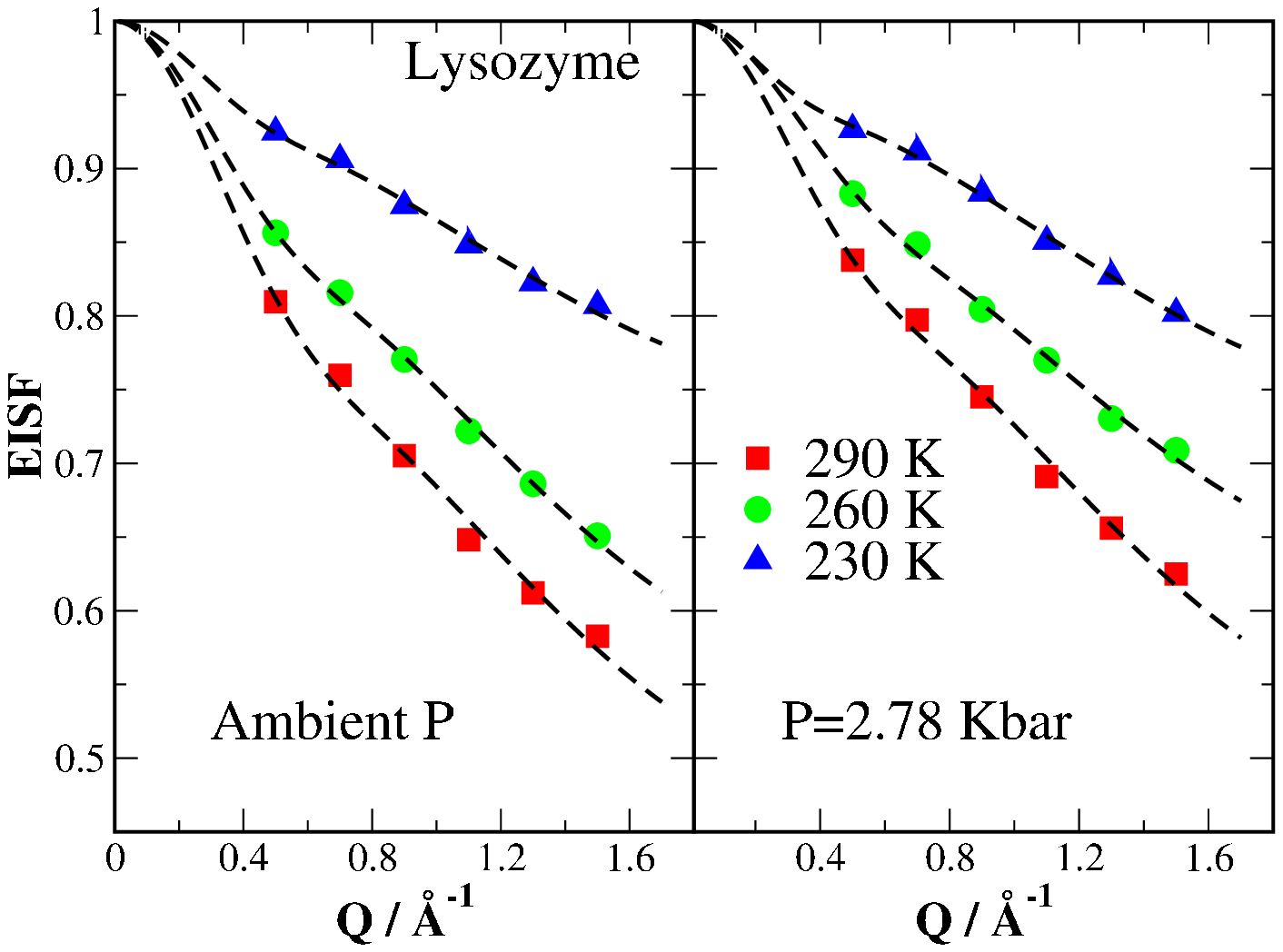}
\caption{Temperature dependence of the elastic incoherent structure factor (EISF) of lysozyme without trehalose: red squares (290 K), green circles (260 K), and blue triangles (230 K). The left panel shows the values at ambient pressure and the right panel indicates those at 2.78 Kbar. The dashed lines are fits of Eq. \ref{eq.eisf} with $a_0$  set to 1.1 {\AA}.}
\label{fig.11}
\end{figure}

\begin{figure}
\includegraphics[width=1.00\linewidth]{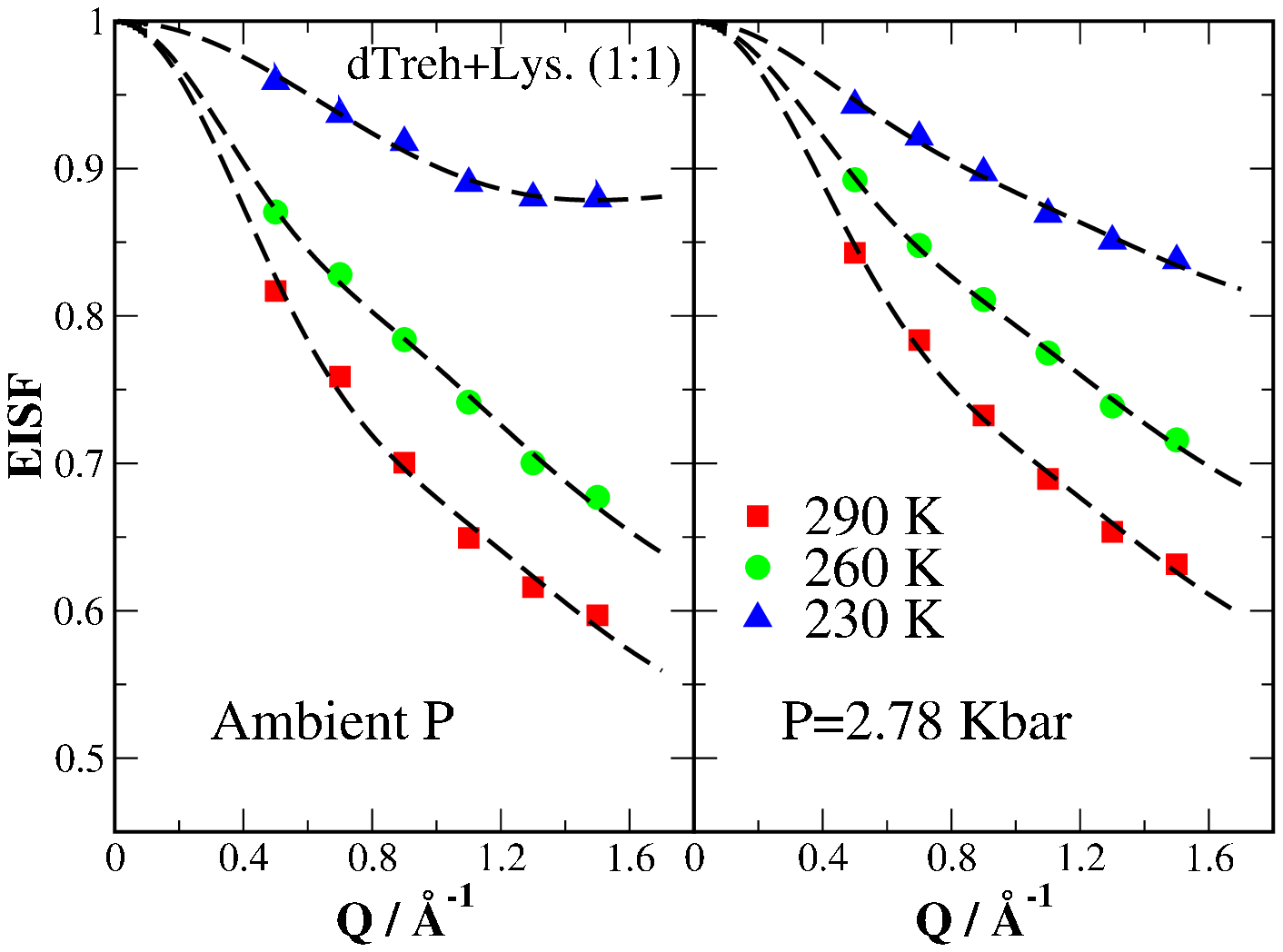}
\caption{Temperature dependence of the elastic incoherent structure factor (EISF) of lysozyme in deuterated trehalose environment.}
\label{fig.12}
\end{figure}

With the confining radii for methyl-groups fixed to its nominal value, fits to the $EISF(Q)$ shown in Figs. \ref{fig.11} and \ref{fig.12} yield an $a_1$ parameter in the range 3-5 {\AA} at all temperatures for both samples. There is no clear systematic dependence of $a_1$ on pressure, within our limited $Q$-range.  We observe a subtle pressure dependence of the population fraction contributing to the rotations of the methyl-groups and those that are not.  These fractions $p_s$ and $f_s$ are summarized in Table \ref{tab1}.

\subsection{Structure Conservation}
\begin{figure}
\includegraphics[width=1.00\linewidth]{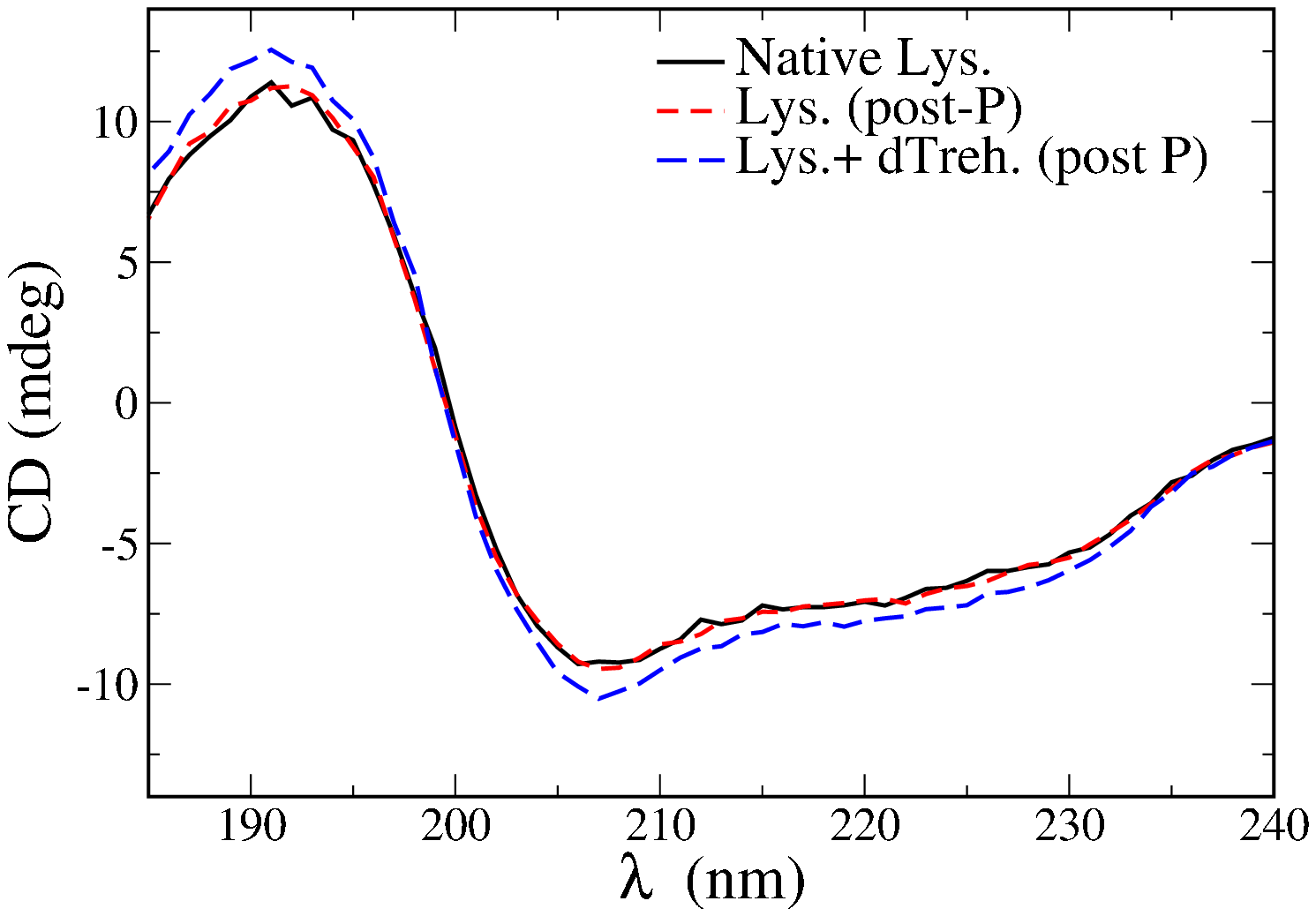}
\caption{Circular dichroism of lysozyme at ambient pressure.  Comparison between a native lysozyme reference sample (black solid line) and samples subjected to high pressure of 2.78 Kbar: Lysozyme alone (red short-dashed line) and lysozyme and trehalose mixture (blue long-dashed line).}
\label{fig.13}
\end{figure}

\begin{figure}
\includegraphics[width=1.00\linewidth]{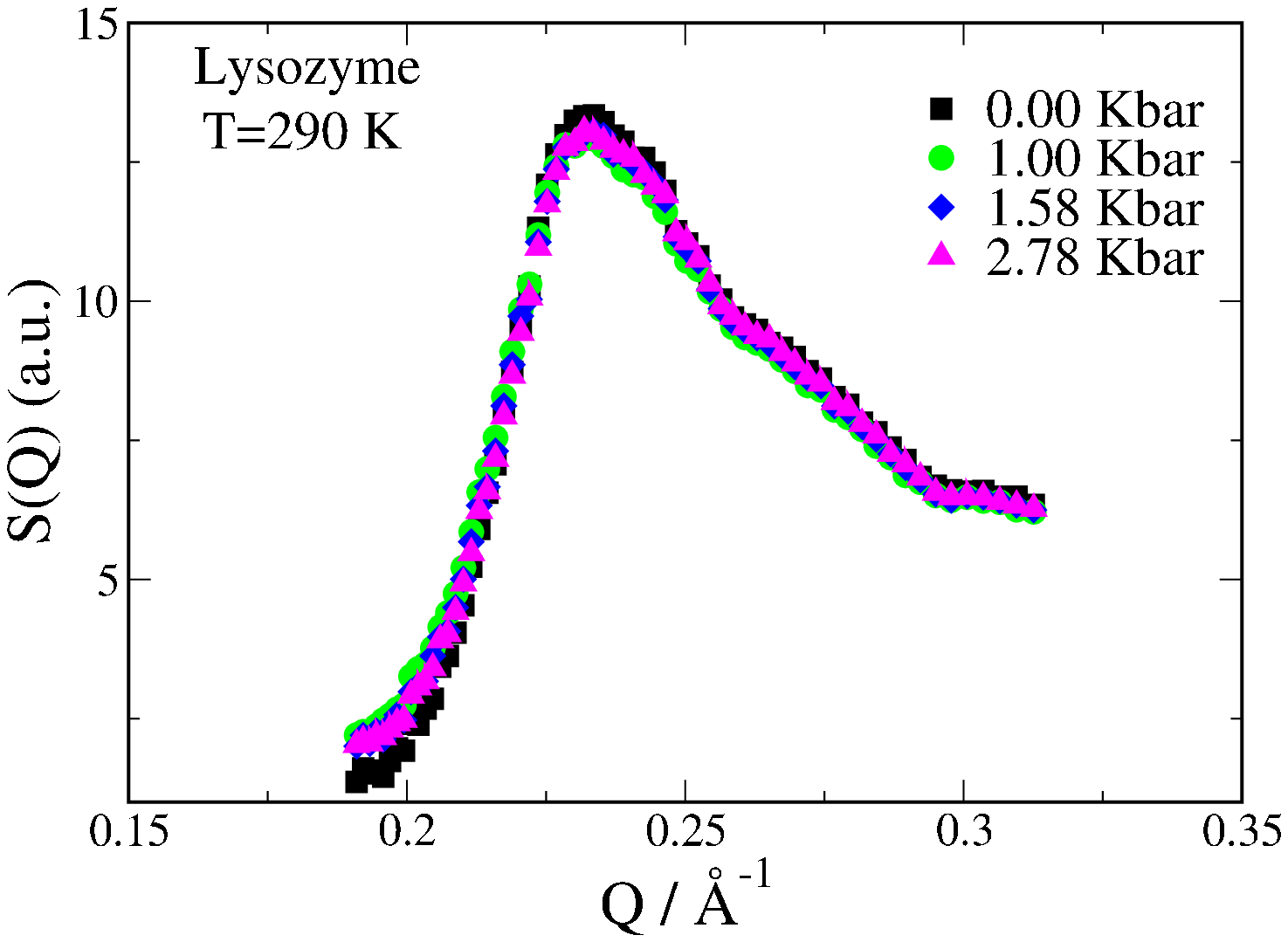}
\caption{Influence of pressure on the observed structure peak of lysozyme at 290 K. Data were taken {\it in-situ} at BASIS using new diffraction detectors. Symbols represent results obtained respectively at ambient pressure (black squares), 1.00 Kbar (green circles), 1.58 Kbar (blue diamonds), and 2.78 Kbar (magenta triangles).}
\label{fig.14}
\end{figure}

Extracting secondary structure contents from circular dichroism spectro-polarimetry (CD) is common practice \cite{Mamontov:10}. We  used CD to check for structure conservation in the protein upon removal of the high pressure.  If structural denaturation or unfolding occurs under pressure, we expect the CD response of the denatured protein to be different than that of the folded unpressurized protein. The measured CD spectrum is the intensity difference between the absorption of  left-handed and right-handed circularly polarized light (in millidegree).  A model spectrum consistent of a linear combination of expected CD responses from $\alpha$-helix and $\beta$-sheets of lysozyme can be used to analyze the data. Based on our CD results summarized in Fig. \ref{fig.13},  it appears that the overall secondary structure of the protein subjected to hydrostatic pressure, is no different than that of the native reference protein, indicating no protein denaturation has occurred under pressure. The minor contrast between the CD intensity of the bare protein and that of the protein-sugar mixture has more to do with the difference in their light absorption coefficients but not with globular structure distortions. This is strengthened by the {\it bonus} low angle diffraction data taken simultaneously during the QENS measurements, which also suggest that the protein maintains its compact and globular structure under pressure. These small angle neutron scattering data collected between 0.18-0.3 {\AA}$^{-1}$ taken {\it in situ} reveal the existence of a secondary structure peak of lysozyme which remains invariant with pressure, as shown by Fig. \ref{fig.14} for D$_2$O-hydrated protein. The peak is centered around 0.225 {\AA}$^{-1}$ for both samples, and corresponds to internal spatial correlations around 28 {\AA}.   It will be interesting to probe much larger length scales, which have been shown to be sensitive to modest pressures \cite{Ortore:09} with small angle scattering, and see what role if any trehalose plays in suppressing denaturation. 

\section{Summary} \label{sec:sum}

The non-reducing disaccharide trehalose is widely known for its thermo-protective benefits for certain micro-organisms and plants in arid regions but its baro-protective properties have yet to be fully demonstrated. In the present study work, we have attempted to understand a consequence of the latter on  the molecular dynamics of lysozyme protein which is submerged in trehalose. The quasi-elastic neutron data indicates no significant slowing-down of the lysozyme dynamics as pressure is increased from ambient pressure to 2.78 Kbar independently of whether or not trehalose was present.   The lack of any observable effect of pressure on the relaxation times  of  lysozyme combined with our CD observations, indicates unambiguously that the overall secondary structure identity of the protein is undisturbed by pressure,  up to at least 2.78 Kbar.  Because of this preservation of the structural integrity of the protein under the pressures probed in the present measurement, we are not able to elucidate the role played by trehalose in baro-protection of bio-molecules. One may be tempted to conclude that perhaps trehalose plays no baro-protection role since it has no particular impact on the QENS signal as pressure is varied, but such conclusion merits  further investigation at the higher pressure where the protein actually unfolds and denatures. 

In spite of this, we have been able to quantitatively document the evolution of the protein dynamics under hydrostatic pressure up to $\simeq$ 3 Kbar, showing no significant changes in the relaxation times with pressure at any temperature.    From model fits to the observed elastic incoherent structure factor or (EISF), we have estimated the molecular fractions of hydrogens in lysozyme that are associated with relevant localized dynamics, as well as the corresponding spatial correlation lengths (3-5 {\AA} ); differentiating between the contributions from the methyl groups, and those of other groups.

It is quite possible that the  relatively low hydration water content used here prohibited the protein from unfolding under pressure, if  denaturation at medium pressures occurs through a solvent mediated mechanism. While this hypothesis can be excluded on the basis of the findings in Ref. \cite{Ortore:09} up to at least 1.5 Kbar, It has also been reported by Hedoux {\it et al.} \cite{Hedoux:11} that the softening of the hydrogen bond network of water due to pressure, could subsequently induces a softer protein dynamics. The key point is that solvent-protein interactions are important, and that protein in solution, would respond differently to hydrostatic pressure than powders would. In fact, Refaee {\it et al.} \cite{Refaee:03} argued using NMR data that buried water molecules play an important role in conformational fluctuation at normal pressures, and are implicated in the nucleation sites for structural changes leading to pressure denaturation or channel opening.

 Future work could explore the effects of pressure on protein dynamics under increased hydration levels -- but more importantly at the higher pressures, in excess of 6-7 Kbar, where Bridgman \cite{Bridgman:014} first reported a denaturation of lysozyme. These pressures are unfortunately not currently achievable by our pressure intensifier, and are also limited by the design of our Al sample holder (rated to $\sim$4.5 Kbar). While  these represent major technical limitations, overcoming them will open up unprecedented opportunities for conducting high pressure research in biological and chemical physics.

\acknowledgments
We thank M. Rucker, S. Elorfi, and M. Loguillo for their excellent technical support with the high pressure equipments, and R. Goyette for his support at the beamline. Much gratitude is due to J. Carmichael for conceiving and designing the high pressure autofrettage holder. We are grateful to W. Heller for critically reading the manuscript. HON and QZ acknowledge the support of the Center for Structural Molecular Biology at ORNL supported by the U.S. DOE, Office of Science, Office of Biological and Environmental Research Project ERKP291. Work at ORNL and SNS is sponsored by the Scientific User Facilities Division, Office of Basic Energy Sciences, US Department of Energy.


\end{document}